\documentstyle[aps,epsf]{revtex}

\begin{document}


\vspace{2cm}

\title{A METHOD TO DETERMINE \\THE IN-AIR
SPATIAL SPREAD OF \\ CLINICAL ELECTRON BEAMS}

\author{M. Vilches, J.C. Zapata, D. Guirado, D. Fern\'andez and
D. Burgos}

\address{Servicio de Radiof\'{\i}sica, Hospital
Universitario ``San Cecilio'', \\
Avda. Dr. Ol\'oriz, 16, E-18012 Granada, Spain.}

\author{A.M. Lallena}

\address{Departamento de F\'{\i}sica Moderna, Universidad de
Granada, \\ E-18071 Granada, Spain.}

\maketitle

\begin{abstract}
We propose and analyze in detail a method to
measure the in-air spatial spread parameter of clinical electron 
beams. Measurements are performed at the center of the beam and below 
the adjustable collimators sited in asymmetrical configuration in order 
to avoid the distortions due to the presence of the applicator.
The main advantage of our procedure lies in the fact that the dose 
profiles are fitted by means of a function which includes,
additionally to the Gaussian step usually considered, a background 
which takes care of the dose produced by different mechanisms that the
Gaussian model does not account for. As a result, the spatial spread is
obtained directly from the fitting procedure and the accuracy permits
a good determination of the angular spread. The way the analysis is 
done is alternative to that followed by the usual methods based on the
evaluation of the penumbra width.
Besides, the spatial spread found shows the quadratic-cubic
dependence with the distance to the source predicted by the
Fermi-Eyges theory. However, the corresponding values obtained for
the scattering power are differing from those quoted by ICRU nr.~35 by
a factor $\sim 2$ or larger, what requires of a more detailed
investigation. 
\end{abstract}


\section{INTRODUCTION}

Nowadays, electron beams have become a tool widely used in radiation
treatment of cancer. However, one of the major difficulties in
the daily clinical procedure is the accurate calculation of the
electron dose distribution. At present, most of the treatment
planning determinations are based on the pencil beam
model,\cite{hog81} which assumes that broad beams are composed
by an infinite number of pencil beams, each one
spreading as predicted by the Fermi-Eyges multiple scattering
theory. In this approach the pencil beams are supposed to
present Gaussian distribution profiles in both space and angular
coordinates, at any point of their trajectory, and then the
corresponding spatial and angular spread parameters are basic 
ingredients to characterize the beams.

The purpose of this work is to determine the spatial spread 
parameter in-air, on the beam axis and for various distances to the 
source. 

The usual techniques to obtain the spatial and angular spreads use 
different relations between the spatial spread and the penumbra width 
(see e.g. Refs.~\cite{hog81,hui87,san88}). However, these procedures 
present a major problem because of the different definitions of the 
penumbra width which can be found in the literature. In 
Refs.~\cite{hog81,kea96}, it is considered to be the average 
spatial separation between the 10 and 90\% isodose levels. The 
ICRU~\cite{icr84} recommended the same definition but for the 20 
and 80\% isodose levels. Finally, in Refs.~\cite{hui87,san88} the 
penumbra width is obtained as the distance between the intersection 
of the tangent at the 50\% point with the 0\%-100\% dose levels. 
In all the cases the measurements refer to a normalized beam 
profile. The main problem is that the values obtained using the
different procedures described above can differ by more than a 30\% 
and therefore the concept of penumbra width is itself misleading. 

Besides, the procedure followed in these works presents additional
error sources which are neither considered nor discussed.
The errors in the measured dose profiles, in 
the determination of the point of 50\% dose and in the calculation
of the corresponding tangent are usually forgotten. All these errors
sum up increasing the indetermination of the penumbra width, with the
consequent ambiguities in the quantities calculated from
it~\cite{zap97}.

A different approach to the problem is followed by
McKenzie~\cite{mck79} and Werner, Khan, and Deibel~\cite{wer82}.
In their work these authors measure the electron dose
distribution behind the edge of a lead block covering a flat
homogeneous phantom in the half plane $x<0$ as in
Ref.~\cite{san88}, but they do not determine the penumbra
width. Instead, they obtain the spatial spread from Gaussian functions 
fitted to the corresponding strip beam profiles. These are calculated 
by deriving the measured dose profiles by means of a two-point 
difference formula. Unfortunately, each step in this method (the 
measurement of the dose profiles, the method to obtain the strip 
profiles and the fit procedure) introduces an error in
the results which is not considered at all. All
these errors propagate to the final results and it is not
possible for the reader to know their accuracy.

In this work we want to address a new procedure to obtain the spatial
spread which, as in Refs.~\cite{san88,mck79,wer82}, is based on
measurements performed at the central area of the beam. However,
our method uses a simple analytical formalism, is plainly
reproducible and shows controlled sources of uncertainty. 

The organization of the paper is as follows. In Sect.~II we
describe the theory underlying the method. Sect.~III is
devoted to the material and methods used in the experimental
side. In Sect.~IV we discuss the results. Finally, we give our
conclusions in Sect.~V.

\section{THEORY}
\label{theo}

As mentioned above, our procedure to determine the in-air
spatial spread is based on the measurement of the dose profiles 
generated by an electron beam below a lead block partially collimating
it. In this section we justify our methodology.

\begin{figure}[ht]
\vspace{-3cm}
\begin{center}                                                                
\leavevmode
\epsfysize = 400pt
\hspace*{.45cm}
\makebox[0cm]{\epsfbox{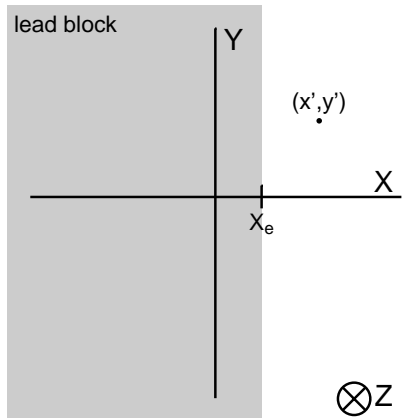}}
\end{center}
\vspace{-5.5cm}
\caption{Experimental setup considered in our work.}
\end{figure}

Let us consider the $xyz$-coordinate system depicted in Fig.~1, to
which the measurements will be referred.  
Let us suppose an infinitely broad beam parallel to the $z$-axis and 
traveling in the direction of increasing $z$. To calculate the dose
deposited by the beam at a given point we use the pencil beam model. In
this approach, it is assumed that broad beams are formed as the sum of
infinite parallel pencil beams, each of them producing dose profiles of
Gaussian type at each $z$. Thus, in our case, the ray centered at the 
point $(x',y')$ in Fig.~1 gives rise to a profile:
\begin{equation}
\displaystyle
D_{(x',y')}(z,x,y) \, = \, D_\infty \, 
\frac{1}{\sqrt{2\pi}\sigma_x(z)} \,\exp 
\left[ -\frac{(x-x')^2}{2\sigma_x^2(z)} \right] \,
\frac{1}{\sqrt{2\pi}\sigma_y(z)} \,\exp 
\left[ -\frac{(y-y')^2}{2\sigma_y^2(z)} \right] \, ,
\end{equation}
where $D_\infty$ is a normalization constant
which actually gives the broad beam electron dose. The
parameters $\sigma_x(z)$ and $\sigma_y(z)$ label the spatial spreads
in the $x$ and $y$ directions, respectively, at a given $z$. 

Let us now assume a semi-infinite lead block partially collimating the
beam, as shown in Fig.~1. We call $X_{\rm e}$ the $x$-coordinate of the
edge of the lead block in our reference system and we
fix the $xy$-plane at $z=0$ coinciding with the lower edge of the
collimator. We can calculate the total dose distribution in $xy$-plane
at a given position $z>0$ as:
\begin{eqnarray}
\nonumber
D(z,x,y) & = & \int_{X_{\rm e}}^\infty {\rm d}x' \,
\int_{-\infty}^\infty {\rm d}y' \, D_{(x',y')}(z,x,y) \\
\nonumber
&= & D_\infty \, \int_{X_{\rm e}}^\infty {\rm d}x' \,
\frac{1}{\sqrt{2\pi}\sigma_x(z)} \,\exp 
\left[ -\frac{(x-x')^2}{2\sigma_x^2(z)} \right] \\
\nonumber
&=& D_\infty \, \int_{-\infty}^x {\rm d}t \,
\frac{1}{\sqrt{2\pi}\sigma_x(z)} \,\exp 
\left[ -\frac{(t-X_{\rm e})^2}{2\sigma_x^2(z)} \right] \\
\label{dose}
&=&  D_\infty \, P(x;X_{\rm e},\sigma_x(z)) \, ,
\end{eqnarray}
where:
\begin{equation}
P(x;m,\sigma)\,=\, \int_{-\infty}^x {\rm d}t \,
\frac{1}{\sqrt{2\pi}\sigma} \,\exp 
\left[ -\frac{(t-m)^2}{2\sigma^2} \right] \, = 
\, \frac{1}{2}\, {\rm erfc}
\left( \frac{m-x}{\sqrt{2}\sigma} \right)
\end{equation}
is the cumulative distribution function corresponding 
to a normal distribution centered at $m$ and with standard 
deviation $\sigma$ and erfc stands for the complementary error 
function~\cite{abr72}.

It is important to note that, the dose distributions given by
Eq.~(\ref{dose}) have the same centroid $X_{\rm e}$, independently of
the $z$-value at which they are measured. This is so because the beam
we have considered up to now is parallel to the $z$-axis of our
reference system.

Eq.~(\ref{dose}) is valid only when the measurement 
plane is irradiated by a uniform, semi-infinite broad beam. In actual 
experiments this is not the case and it is necessary to consider the 
corresponding corrections.

First of all, it is obvious that the actual beam is finite and, as a
consequence, it exhibits a physical end in the open (not collimated)
area. This produces a certain distortion in the dose profiles which
will differ from the Gaussian shape expected for infinitely broad
beams. In order to minimize these differences, we focus our attention
on the data acquired below the lead collimator. Therein, we expect to 
be sufficiently far away from this physical end of the beam and we can
assume that the Gaussian approach is enough reasonable to describe the
profiles.

On the other hand, the finite dimensions of the source, together with the
fact that the distance from the source to the measuring plane is not
infinity, make the actual beam to diverge. This gives rise 
(see e.g. Ref.~\cite{hui87}) to two effects. The
first one is that the constant $D_\infty$ in Eq.~(\ref{dose})
depends on the $z$-value. Then the dose at a given point is:
\begin{equation}
\label{realdose}
D(z,x,y) \, = \, D_\infty(z) \, P(x;X_{\rm e},\sigma_x(z)) \, .
\end{equation} 
For electron beams generated in linear accelerators (LINAC), it is
possible, in practice, to define a point virtual source. As a
consequence, the dependence of $D_\infty$ with $z$ must verify the
well known inverse squared law. We will check this point {\it a 
posteriori}\/ directly on the measured profiles (see
Subsect.~\ref{anaD}).

A second effect of the divergence of the beam in the actual experiment
is that, in general, the centroid of the dose distributions will vary
with the $z$-coordinate of the measuring plane~\cite{hui87}. Then,
the dose profiles will be given by:
\begin{equation}
\label{actualdose}
D(z,x,y) \, = \, D_\infty(z) \, P(x;x_{\rm cent}(z),\sigma_x(z)) \, ,
\end{equation}
where $x_{\rm cent}(z)$ represents, for each $z$ value, the $x$ 
position of the centroid of the distribution. 

It is worth to point out that only if the $x$-position of
the point virtual source of the beam is exactly $X_{\rm e}$, the
centroids $x_{\rm cent}(z)$ would be independent of $z$. In such a case
these centroids will equal $X_{\rm e}$. It is evident that 
this situation cannot be ensured in actual experiments; nevertheless,
this particular case is included in the general equation
(\ref{actualdose}). 

Geometrically, the centroid $x_{\rm cent}(z)$ can be understood as 
the radiological projection of the edge of the collimator produced by 
the beam and it is expected to behave linearly 
with $z$ (see Ref.~\cite{hui87}). An immediate result is that the 
equation:
\begin{equation}
\label{cent-z0} 
x_{\rm cent}(z=0) \, = \, X_{\rm e}
\end{equation}
must be verified. In our experimental procedure (see
Subsect.~\ref{centroid}) we measure these positions $x_{\rm cent}(z)$
and we check that they show the required linear behavior and that
Eq.~(\ref{cent-z0}) is satisfied.
 
Finally, it is worth to note that, in the actual experiments,
measurements are performed by means of an ionization chamber sited in
an electro mechanical device which permits the positioning of the
chamber. Then, the source-plus-collimator system will not be perfectly
aligned, in general, with the coordinate system in which measurements 
are done. One can expect
that this new system is both shifted and rotated with
respect to the measurement system and this must be taken into
account. However,
these effects can be minimized by controlling with the standard
procedures (optical, mechanical, etc.) the positioning of the gantry
head of the accelerator and we will assume they are incorporated in
the values of $x_{\rm cent}(z)$ we determine experimentally.

In view of the previous discussion, the fitting function we adopt to
analyze the experimental dose distributions is the following:
\begin{equation}
\label{fitf}
D_{\rm fit}(z,x) \, = \, D_\infty(z) \, 
P(x;x_{\rm cent}(z),\sigma_x(z)) \,  + \, B(z,x) \, .
\end{equation}

We have add the ``background'' function $B(z,x)$ 
in order to take care of different
contributions to the dose which the Gaussian model does not account
for. Thus, part of the dose due to the bremsstrahlung and 
the dose due to electrons scattered at the gantry head, at
the measurement device and its surroundings, as well as those
scattered in-air with large angles, are supposed to be described by 
this function. The particular functional dependence of $B(z,x)$ with 
$x$ will be considered in Subsect.~\ref{sect:fit} and its role in the
model we propose will be discussed in Subsect.~\ref{anaB}. As we
quote below, the contribution of the background is not at all 
negligible and its consideration in this simple way permits to obtain 
very good fits of the measured profiles.

\section{EXPERIMENTAL SETUP}
\label{sect:exp}

As mentioned above, our interest is to determine the in-air spatial 
spread of clinical electron beams produced by a 
linear accelerators. Also, we want to investigate its dependence with
the distance to the lower edge of a lead block (actually, the inner
jaw) collimating the beam.
To do that, we have measured the corresponding relative dose profiles 
in air by using a Wellh\"offer WP-700 system which incorporates an 
electrometer Wellh\"offer WP-5007 and an ionization chamber 
Wellh\"offer IC-10 of 0.12~cm$^3$ sited in an electro mechanical 
device which permits the tridimensional positioning of the chamber 
with a theoretical precision of 0.1~mm. The dose profiles have been 
acquired in continuous mode, by displacing the chamber in the
$x$-axis direction (at $y$=0) and for $z$ values ranging 
between 50 and 80~cm. Beams with nominal energies varying from 6 to 18~MeV 
generated by a Siemens Mevatron KDS have been considered. This ensures
the generality of the results for similar machines. 
All the profiles obtained have been treated by means of the software
package WP-700 Version 3.20.02 accompanying the measurement system.

In our reference system ($z=0$ at the bottom edge of the jaws), the 
isocenter lies at $z=73$~cm and the scattering foils are sited at
$z=-27$~cm. 

The reproduction of the conditions established in the theoretical 
hypothesis was achieved by collimating the beam to the central axis 
using the adjustable collimators of the LINAC in asymmetric
configuration. This point deserves a comment. Electron dose 
computation programs require as input data the spatial and/or angular
spreads of the beam. The usual practice is to determine these
parameters below the applicator at treatment distances. In this way,
the obtained values are considered to be clinically relevant because
the possible modifications in the parameters introduced by the
presence of the applicator are taken into account. However, our
interest stands for a more basic question. As mentioned in the
Introduction, what we want to do is to determine the spatial spread of
the beam in order to have more information to characterize it. In this
sense we are interested in eliminate all the possibles distortions
induced by the use of the different clinical devices. Besides we try
in this paper to show the possibilities of the analysis technique we 
propose and to study its feasibility. We think that the experimental
setup we consider provides the cleanest experimental situation to
complete our purposes. We leave for a following work the investigation
of the behavior of the parameters of interest when applicator and
phantom are introduced~\cite{zap97}.

\subsection{Determination of the position of the centroid, 
$x_{cent}$}
\label{centroid}

As we have previously discussed, the position of the 
centroid of the dose distribution shifts in the transverse direction 
when the beam goes downstream. The procedure we have performed to 
determined the values $x_{\rm cent}(z)$ at each $z$ is based in a
series of measurements done with the following experimental setups:

\begin{description}
\item{\sc Setup~1.}
First, we have moved the right collimator to the center (position 
$X_{\rm e}$) maintaining the left one apart. In this situation we have
obtained four dose profiles, without the reference chamber, for each
energy and for each of the four values of $z$ selected: 50, 60, 70 and
80 cm. Each profile has been taken crossplane (with $y=0$) and varying
$x$ from -2 to 2 cm.

\item{\sc Setup~2.}
Next, we have moved the left collimator to the center. In order to 
check that beams are completely collimated we have measured 
four dose profiles for the largest energy (18~MeV) at $z=50$~cm. We 
have checked that the dose level in this situation is zero. This 
ensures that the left collimator is at $X_{\rm e}$ also.

\item{\sc Setup~3.} 
Finally, we have moved the right collimator apart. With this setup we 
have got four dose profiles, again without the reference chamber, for 
each energy and for the same values of $z$ as in Setup~1. 
\end{description}

The role of the reference chamber in our procedure deserve some
additional explanations. It is obvious that the way we determine the 
$x_{\rm cent}$ values imposes the necessity of guaranteeing an equal 
normalization for the profiles obtained in the two asymmetrical
configurations corresponding to Setups 1 and 3. However, it has not
been possible to situate the reference chamber in a manner that it 
were irradiated identically in both setups. This is the reason why we
switched off it when the profiles used to measure $x_{\rm cent}$ were
taken. 

The experimental procedure followed (which ensures the absence of 
radiation when the two collimators are closed in Setup~2) allows us to 
obtain $x_{\rm cent}$ by finding the cross point between couples of
the symmetrical dose profiles measured with Setups~1 and 3. To
determine the cross point of a given pair we have, first, calculated
the difference between the two profiles of the couple, then found the
zero of the resulting function and, finally, checked that the two
profiles are symmetric with respect to the cross point. The
manipulation of the profiles as described here has been done with the 
software package WP-700 of the measurement system.
The four profiles measured with each one of the Setups 1 and 3,
provide us with 16 experimental values of $x_{\rm cent}$ (for each 
energy and $z$). These values give a sufficient 
statistics and permit us to calculate the corresponding mean, 
$\overline{x}_{\rm cent}$, and standard deviation, 
$\sigma_{{x}_{\rm cent}}$, values. The results obtained in this way are
shown in Fig.~2, as a function of $z$ (in cm), for the five energies 
considered.\footnote{Throughout this work uncertainties are given with
a coverage factor $k=1$~\cite{iso92}}

Here, a particular detail needs a comment. It is obvious that with
the two asymmetric setups we use to determine the centroids, the shadow
is cast by a different part of the collimator: when {\sc Setup 1} is
used, the top of the jaw forms the shadow, while it is the bottom of
it which is casting the shadow in {\sc Setup 3}. However, the
description of how the shadow is formed is not as easy as this vision
implies. In fact, it has been shown \cite{san88} that the shadow is
cast by the full collimator and only when the angles subtended by the
source are bigger than 2$^{\rm o}$ the effect of the edges of the
collimator is important. Taking into account that our jaws are 6~cm
thick and the positions derived for the virtual sources, a simple
geometrical calculation tells us that, in our case, these angles are 
of a few tenth of degree at most. Then the effect due to the
differences in the cast of the beam in both asymmetrical setups are
smaller than the spread we obtain for the centroid in our statistics
and can be neglected.

\begin{figure}[b]
\vspace*{-0.1cm}
\begin{center}                                                                
\leavevmode
\epsfysize = 400pt
\hspace*{.45cm}
\makebox[0cm]{\epsfbox{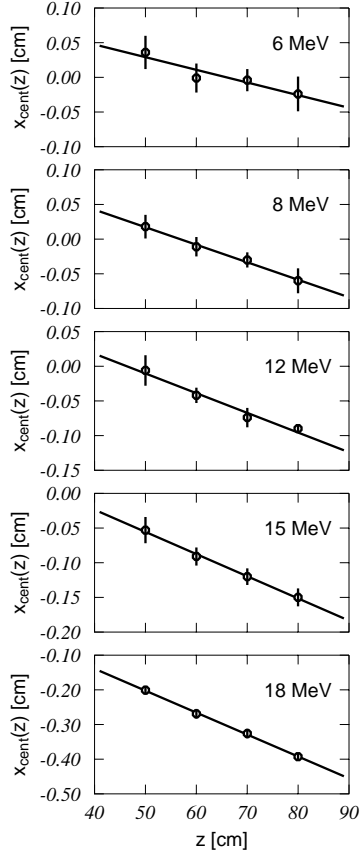}}
\end{center}
\vspace{-2.7cm}
\caption{Experimental values of $x_{\rm cent}$, obtained in
our work, as a function of the distance $z$. The straight lines 
correspond to the linear regression of the data.}
\end{figure}

\vspace{-.2cm}

\begin{table}
\caption{Results of the fits of the values
$\overline{x}_{\rm cent} \pm 
\sigma_{x_{\rm cent}}$ obtained in our experiment to a linear 
function of the
type $\overline{x}_{\rm cent}\,=\,a\,*\,z\,+\,b$. Note that $b$
defines the position $X_{\rm e}$ of the edge of the collimator.} 
\begin{tabular}{ccccccc}
Energy [MeV] &~~~~& a &~~~~& b [cm] &~~~~& Correlation \\ \hline
~6 && -0.0017 $\pm$ 0.0009 && 0.11 $\pm$ 0.06 && -0.946 \\
~8 && -0.0025 $\pm$ 0.0007 && 0.14 $\pm$ 0.04 && -0.997 \\
12 && -0.0025 $\pm$ 0.0004 && 0.11 $\pm$ 0.03 && -0.988 \\
15 && -0.0032 $\pm$ 0.0006 && 0.10 $\pm$ 0.04 && -0.998 \\
18 && -0.0063 $\pm$ 0.0005 && 0.11 $\pm$ 0.03 && -0.999 \\
\end{tabular}
\end{table}

Once the transverse shifts of the beam edge offset have been measured,
we want to check if their variation with $z$ is linear or not. In 
Table~I we give the values of the parameters of the linear regression
performed for the experimental centroids. The obtained fits are also 
shown in Fig.~2 (straight lines). As we can see the correlation
coefficients indicate that the way we have introduced in our model the
beam divergence is in agreement with the experimental findings, at
least in what refers to this aspect of the centroid shift.

Also, it is worth to point out two details concerning the results of
this part of the experiment. First, it is satisfactory that the 
parameter $b$ appears to be constant with the energy. This parameter
is measuring (see Eq.~(\ref{cent-z0}))
the offset at the collimating plane ($z=0$) and it is
obvious that it should be the same in all cases, because the position 
of the collimator does not change. This is showing again that the
procedure of analysis we are carrying out is robust and correct. 

Second, the fact that $a$ varies with the energy is due to the 
differences in the focal spots at the primary foils for each energy. 
The variation of these focal spots found using the regression we have 
obtained is of 1.12~mm, which is compatible with the technical
specifications of our accelerator.

\subsection{Determination of the dose profiles data and their errors}

The second point relevant in the experimental part corresponds to the
way we have obtained the profiles to be fitted and how the
corresponding errors have been estimated. 

It is obvious that, due to the character of the process itself, there
exists a statistical uncertainty in the dose profiles measured. In
order to take care of this point we have measure five new profiles
with Setup~3, now with reference chamber, for each energy and for 
seven values of $z$: 50, 55, 60, 65, 70, 75 and 80 cm. These
measurements have been performed together with those used to determine
the positions $x_{\rm cent}$, what ensures the validity of the
results obtained in the previous subsection to analyze these new 
profiles. The reason to take these new profiles with the reference
chamber (contrary to what we have done for the profiles used to
determine the centroid positions) is that it permits a better
statistics and a quick procedure,
with the obvious beam time saving.

The five dose distributions corresponding to each energy and
$z$ are then processed in order to generate the data. First, we have
sampled them at different $x$ positions. In this respect it is worth
to say that, as we mentioned in Sect.~\ref{theo}, we have
considered the data obtained below the collimator in order to 
fulfill the requirement of being at sufficient distance from the
physical end of the beam to avoid the possible distortions.
Thus, only the values of the dose up to $x=0$ are taken into account.
In principle we should have taken into account the doses up to
$x_{\rm cent}$ to insure we use only the data in the shadow of the
collimator. However, when the profiles were taken the positions of the
centroids for the different energies and $z$-values were not known and
we sampled up to $x=0$. This means only a few millimeters
around the actual values of $x_{\rm cent}$ and we assume this does not
invalidate our assumptions. The data obtained in the sampling are 
averaged to obtain the $\overline{D}(x)$ values at each $x$. 

The second step is to estimate the errors accompanying these dose
data. To do that we have followed the prescriptions of 
Ref.~\cite{iso92}. 
In our case we have two error sources. First, there exists the obvious
uncertainty associated to the statistical behavior we have just 
considered. This kind of error can be evaluated by means of the standard 
deviation, $\sigma_D(x)$, we obtain simultaneously to the mean values 
$\overline{D}(x)$. This is so because measuring in continuum mode, as 
we have done, implies the simultaneous consideration of the
statistical variation of both the dose and the positioning of the 
measuring chamber. The corresponding uncertainty is treated as of type
A (that is, not assuming an {\it a priori} distribution) in the 
nomenclature of Ref.~\cite{iso92} and then it is evaluated by 
considering directly the observed distribution. In our case the values
found for the relative error corresponding to this statistical spread 
is below 1\% in all cases. (Here the \% refers to the relative units in
which the profiles are measured).

The second source of error is that linked to the precision of the data
acquisition system. This is basically different to the first one and
following Ref.~\cite{iso92} is classified as of type B. To evaluate it,
we have considered it is the one corresponding to a digital 
measurement apparatus which is~\cite{iso92}:
\begin{equation}
\displaystyle
\Delta_D \, = \, \frac{1}{\sqrt{3}} \, \delta_D \, = \, 0.029\%,
\end{equation}
where in our case $\delta_D = 0.05\%$ is half of the nominal
precision of the measurement device.

The total error is then calculated as the quadratic sum of both errors:
\begin{equation}
 \Sigma_D(x) \, = \, \sqrt{ \Delta_D ^2 + \sigma_D ^2(x) } \, .
\end{equation}
Our data are the relative dose for each energy, $z$ and position $x$,
which are given by $\overline{D}(x)\pm \Sigma_D(x)$. These values have
been fitted as described below.

\section{Results}

In what follows we discuss the results we have obtained in the analysis
of the dose profiles measured as described above. 

\subsection{Fitting procedure}
\label{sect:fit}

The dose profiles obtained from our measurements have been fitted using
the fitting function in Eq.~(\ref{fitf}) and by means of the 
Levenberg-Marquardt method~\cite{pre92}. As indicated in
Sect.~\ref{theo}, in order to complete the model, we must define
the functional dependence of the background function $B(z,x)$ with the
position $x$. We have assumed the simplest 
possibility, that is, the function $B(z,x)$ is constant at each $z$,
$B(z,x)\equiv B(z)$. In any case, we have checked that the use
of other functional dependences (that is, $B(z,x)$ linear or 
quadratic, in $x$) does not change the conclusions obtained with 
respect to the spatial spread. 

With our choice for the background function, the fit provides the 
quantities $D_\infty(z)$, $\sigma_x(z)$ and $B(z)$ once an input value
of $x_{\rm cent}$ is given. In our experiment, $x_{\rm cent}$ was not 
measured for the intermediate $z$ values 55, 65 and 75~cm.  
Then, in order to be fully consistent in the fitting procedure, 
we have used the values of $x_{\rm cent}$ obtained from the 
linear regression quoted in Table~I and shown in Fig.~2. In any case 
we have checked that same results are obtained in the fits 
when the experimental $x_{\rm cent}$ values are used in those cases 
where they are available.

The fact that $D_{\rm fit}$ as given by Eq.~(\ref{fitf}) is not linear,
does not allow an simple analytical estimation of the propagation 
of the error of $x_{\rm cent}$. Thus, to evaluate the errors 
corresponding to the three quantities defining our fitting function,
we have developed a procedure of Monte Carlo type, 
which is based on the following steps:
\begin{enumerate}
\item From each original profile, we have generated a set of data,
which is built by random values normally distributed around the 
original ones and with standard deviations equal to the associated 
errors. 
\item We have also generated a random value for $x_{\rm cent}$ 
considering the normal distribution obtained in the linear regression
of the experimental values quoted in Subsect.~\ref{centroid}.
\item We have fitted the resulting profile with the function 
(\ref{fitf}) again by means of the Levenberg-Marquardt method.
\end{enumerate}
The repetition of this three step procedure provides a set of values 
for each one of $D_\infty(z)$, $\sigma_x(z)$ and $B(z)$
and then it is possible to evaluate the corresponding standard deviations
for these three quantities. The procedure is repeated until the convergence 
in the standard deviation values is achieved. We have checked that this 
happens for a number of random generations between 5000 and 10000. All the 
results we discuss in the following have been obtained with 10000 
generations. The final values found for the standard deviations
are taken to be the errors in the three parameters resulting from the fit. 

In what follows we discuss the results obtained in the procedure we
have followed to fit the relative dose profiles.

\subsection{Analysis of $D_\infty(z)$}
\label{anaD}

One of the consequences of considering the beam divergence in our
model is the fact that the dependence of $D_\infty(z)$ with $z$ should
follow the inverse square law. In order to check if this is the case, 
we show in Table~II the parameters $a$ and $b$ of the linear 
regressions $[D_\infty(z)]^{-1/2}\,=\,a\,*\,z\,+\,b$ obtained
directly from the values of $D_\infty(z)$ found in the fitting
procedure. As we can see, the data behave as expected (see the
correlation factor quoted in the table), what ensures the 
consistency of the method we have followed. The important point here 
is that we have not measured directly $D_\infty(z)$ but we have used 
the values obtained from the fit of the tails of the dose profiles 
below the lead block. 

\vspace{.3cm}

\begin{table}[t]
\caption{Results of the linear regression of the values 
$[D_\infty(z)]^{-1/2}$ obtained in our fitting procedure with a fitting
function of the type $[D_\infty(z)]^{-1/2}\,=\,a\,*\,z\,+\,b$. Also the
position of the virtual effective source, $z_{\rm vir}=-b/a$, is
given.} 
\begin{tabular}{ccccccccc}
Energy [MeV] &~~~& a [cm$^{-1}$] &~~~& b &~~~& 
Correlation  &~~~& $z_{\rm vir}$ [cm] \\ \hline
~6 && (~8.39 $\pm$ 0.05)$\times 10^{-4}$ && (2.09 $\pm$ 0.03)$\times 10^{-2}$ 
&&  0.99998 && -24.8 $\pm$ 0.5 \\
~8 && (~9.83 $\pm$ 0.05)$\times 10^{-4}$ && (2.29 $\pm$ 0.03)$\times 10^{-2}$ 
&&  0.99968 && -23.1 $\pm$ 0.5 \\
12 && (10.65 $\pm$ 0.05)$\times 10^{-4}$ && (2.42 $\pm$ 0.04)$\times 10^{-2}$
&&  0.99998 && -23.2 $\pm$ 0.5 \\
15 && (11.22 $\pm$ 0.08)$\times 10^{-4}$ && (2.52 $\pm$ 0.05)$\times 10^{-2}$ 
&&  0.99998 && -23.4 $\pm$ 0.7 \\
18 && (12.02 $\pm$ 0.10)$\times 10^{-4}$ && (2.84 $\pm$ 0.06)$\times 10^{-2}$ 
&&  0.99998 && -23.6 $\pm$ 0.7 \\
\end{tabular}
\end{table}

One additional point of interest is that these linear regressions 
allow us to calculate the position of the virtual point source, 
$z_{\rm vir}=-b/a$. 
The values obtained are also included in Table~II, and, as we can see,
the $z_{\rm vir}$ are, for the five energies considered, in a range 
smaller than 2~cm. 

The case of 6~MeV deserve a particular comment. As we can see, 
the result obtained for this energy is the one showing a larger 
deviation and, in fact, it differs in 1~cm roughly with respect to 
the mean of the values of the remaining energies. This is due to the
fact that, contrary to what happens for other energies, no primary foil 
is present in the case of 6~MeV beams. Thus, a shift in $z_{\rm vir}$
of the order of the width of the foil carrier (which is around 1~cm) must 
be expected in this case. Then, our method is able to reflect in a
direct way the architecture of the accelerator head and this
shows up again its precision.

In what follows we normalize, for each energy, the dose values to the 
values $D_\infty(z=0)$ obtained from the regressions given in Table~II
for $[D_\infty(z)]^{-1/2}$. This affects the experimental dose data
and the values of the parameters $D_\infty$ and $B(z)$ shown
below. Obviously, the values of $\sigma_x(z)$ are not modified by this
renormalization.

\subsection{Analysis of $B(z)$}
\label{anaB}

The second aspect concerning the results of the fits of the dose
profiles we are interested in discussing is the role of the background
function $B(z)$. As previously indicated, the reason to include it in
our model is to take into account the different scattering processes
(part of the dose due to the bremsstrahlung and 
the dose due to electrons scattered at the gantry head, at
the measurement device and its surroundings, as well as those
scattered in-air with large angles) 
which are not described by the pure Gaussian profiles
and which, nevertheless, are important to achieve a good description of the
experimental data. In Fig.~3 we show the variation with $z$ of the values of
$B(z)$ (normalized as indicated above), obtained from the fit, for the five 
energies we are considering. As we can see, the general trend is, in
all cases, to
reduce its strength with increasing $z$. This indicates that the
electrons scattered in the measuring device seems to be almost
negligible because their contribution should be more relevant at 
large distances, where the bottom of the electro mechanical device for 
the positioning of the ionization chamber is closer to it.
In any case, to establish which scattering mechanism is the main
responsible for this background function is not an easy task, mainly
because one can expect that, as it occurs for the Gaussian part of the
profiles, various processes contribute to this background dose.

Despite the fact that the values of $B(z)$ 
are small, it is worth to point out that its inclusion in
the model is crucial for the good description of the data. This can be
seen in Fig.~4, where we show the results obtained for the two
extremal energies we consider (6 and 18~MeV) and for the smallest and 
largest $z$ positions. Therein solid curves correspond to the fits
performed with our model, while dashed curves represent the best fits
of the data obtained with a pure Gaussian profile (that is with
$B(z)=0$ in our model). In both cases the values of $x_{\rm cent}$
obtained experimentally as discussed above have been used. Besides,
both calculations, as well as the data, have been normalized to the
values $D_\infty(z=0)$ obtained within our model. In
Table~III we give the values found for the parameters in these fits.
The first point to note is the goodness of the fits produced by our
method. On the other hand, it is worth to point out that the values
obtained for the pure Gaussian model differ clearly from 
those obtained within
our approach (including the corresponding uncertainties.) 

\subsection{Analysis of the spatial spread, $\sigma_x(z)$}

Our main interest in this work concerns with the determination of the
spatial spread of clinical electron beams in order to characterize them.
In our model, the spatial spread determined in the fitting procedure 
corresponds to the Gaussian part of the dose profiles measured. 
According to Fermi-Eyges theory, this spatial spread must show a 
quadratic-cubic dependence with $z$. In particular, this dependence 
is given by (see e.g. Ref.~\cite{icr84}):

\begin{equation}
\label{fits2}
\displaystyle
\sigma_x^2 (z) \, = \, \sigma_{\theta,{\rm i}}^2 \, z^2 \, + 
\, \frac{1}{6} \, T(E) \, z^3 \, ,
\end{equation}
where $\sigma_{\theta,{\rm i}}^2$ is the initial quadratic angular 
spread and $T(E)$ is the air linear scattering power corresponding to 
the energy $E$ and we assume it to be independent of $z$.
Both parameters can be found by fitting with Eq.~(\ref{fits2}) the 
$\sigma_x(z)$ values obtained from the fitting of the dose
profiles. Instead of doing directly this, and in order to simplify 
the analysis, we have linearly regressed the quantity
$\sigma_x^2 (z)/z^2$ as a function of $z$. The values obtained are
given in Table~IV. The regression lines we have found in this 
procedure are plotted in Fig.~5 with full lines. It is apparent the
good agreement produced in this case with the experimental data, what
guarantees the feasibility of the previous assumption concerning the
significance of the spatial spread we have determined.

\newpage

\begin{figure}
\vspace*{-3.5cm}
\begin{center}                                                                
\leavevmode
\epsfysize = 400pt
\hspace*{.45cm}
\makebox[0cm]{\epsfbox{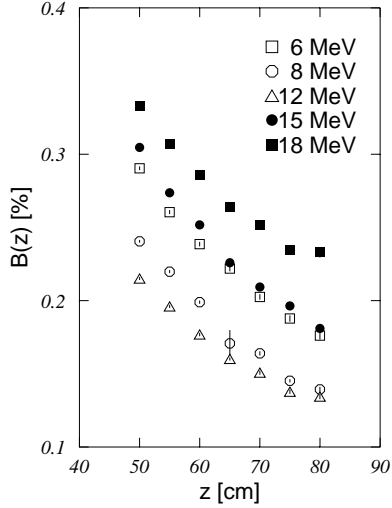}}
\end{center}
\vspace{-4.5cm}
\caption{Values of $B(z)$ obtained in the 
fitting procedure and normalized to $D_\infty(z=0)$ as described in
the text.}
\end{figure}

\begin{figure}[h]
\vspace*{-4.5cm}
\begin{center}                                                                
\leavevmode
\epsfysize = 400pt
\hspace*{.45cm}
\makebox[0cm]{\epsfbox{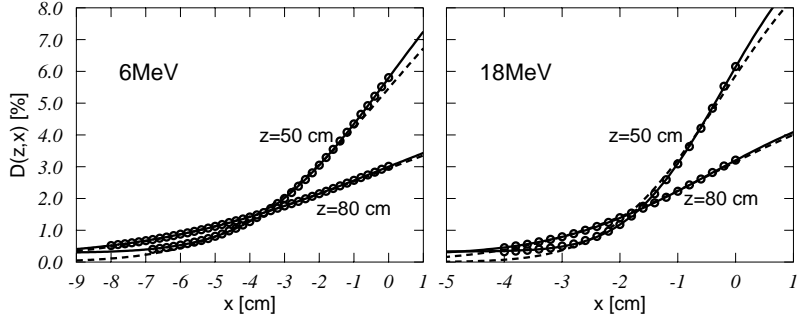}}
\end{center}
\vspace{-5.5cm}
\caption{Experimental dose profiles for 6 (upper panel) and 
18~MeV (lower panel).
Solid curves have been obtained by fitting the experimental doses 
with the function of Eq.~(\protect\ref{fitf}). Dashed lines correspond
to the best fits obtained with pure Gaussian profiles ($B(z)=0$).
The values of the parameters $D_\infty(z)$, $\sigma_x(z)$ and $B(z)$ found 
in both fits are given in Table~III. In both cases the experimental
values of $x_{\rm cent}$ have been considered. At the scale of the figure,
the errors bars are smaller than the symbols used to represent the data.
The normalization to $D_\infty(z=0)$ as described in the text is included.}
\end{figure}

\vspace{.5cm}

\begin{table}
\caption{Values of the parameters 
$D_\infty(z)$, $\sigma_x(z)$ and $B(z)$ obtained in the fits of the 
experimental dose
profiles measured for the extremal energies considered in our work (6
and 18~MeV) and the shortest and largest distances to the source (50
and 80~cm). The values quoted as ``Our model'' have been found
with Eq.~(\protect\ref{fitf}), while in those quoted as ``Pure Gaussian''
the background function $B(z)$ has been taken to be zero. The values
of $D_\infty(z)$, for both models, and $B(z)$ are normalized as
described in the text.}
\begin{tabular}{ccccccccccc}
  & &~~~~& \multicolumn{5}{c}{``Our model''} &~~~& 
\multicolumn{2}{c}{``Pure Gaussian''} \\
\hline
Nominal Energy & $z$  && $D_\infty(z)$ &~~& 
$\sigma_x(z)$ &~~& $B(z)$ & &
$D_\infty(z)$ & $\sigma_x(z)$ \\
\protect[MeV] & [cm]& & \protect[\%] && [cm] && [\%] && [\%] & [cm] \\ \hline
~6 & 50 && 11.08$\pm$0.05 && 2.99$\pm$0.01 && 0.289$\pm$0.003 && 
 11.01 & 3.5 \\  
   & 80 && ~5.65$\pm$0.02 && 5.19$\pm$0.02 && 0.171$\pm$0.007 && 
 ~5.87 & 5.7 \\ 
\hline
18 & 50 && 10.31$\pm$0.07 && 1.30$\pm$0.01 && 0.325$\pm$0.002 && 
 10.64 & 1.5 \\
   & 80 && ~5.20$\pm$0.02 && 2.12$\pm$0.01 && 0.222$\pm$0.06 && 
 ~5.59 & 2.4 \\
\end{tabular}
\end{table}

\begin{figure}[ht]
\vspace*{-3.5cm}
\begin{center}                                                                
\leavevmode
\epsfysize = 400pt
\hspace*{.45cm}
\makebox[0cm]{\epsfbox{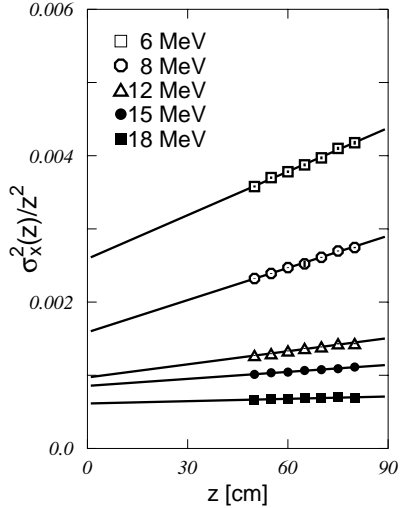}}
\end{center}
\vspace{-4.3cm}
\caption{Values of $\sigma_x^2(z)/z^2$ obtained in the 
fitting procedure of the dose profiles. Full lines correspond to the 
fits performed with the function of Eq.~(\protect\ref{fits2}). 
The error bars are smaller than the width of the symbols used.}
\end{figure}

\vspace{1cm}

Scattering powers are important because they permit to find the angular
spread at any $z$. In fact, $\sigma_{\theta}^2 (z)$ behaves linearly
with $z$ and the slope is precisely $T(E)$. Then it is relevant to
compare the values obtained in our approach with those given in other
references. To do that the first point to elucidate is to fix the beam
energy at which the scattering power is determined. The beam energies 
we have quoted up to now are actually the nominal ones. To calculate
$T(E)$ we have used, instead, the mean beam energies at the isocenter.
Following the prescription of Ref.~\cite{icr84}, these energies have been
obtained from the equation $\overline{E}_0 \,=\,C_6 \, R_{50}$, where 
$C_6=2.33$~MeV/cm. The parameter $R_{50}$ is the half-value depth,
which is determined by range measurements in water at 
SSD=100~cm and with broad beam. The values obtained, $\overline{E}_0$,
are also given in Table~IV. With these values of the mean energies
corresponding to our beam, we have calculated the values of $T(E)$ 
by interpolating those quoted in Table~2.6 of Ref.~\cite{icr84} 
and correcting them for M{\o}ller scattering.

As one can see, these values differ from those obtained with our model
by a factor $\sim 2$ or even bigger. In principle, this means that
the lineal scattering powers we have found suggest energies for the beam 
considerably bigger than the mean energies determined experimentally 
as mentioned above. However one should be careful with the
interpretation of these results because of the approach considered
in Ref.~\cite{icr84} to obtain the $T(E)$ values is quite different
from ours. In this sense, a more detailed investigation in this 
direction is needed before establishing definitive conclusions. 
The possibilities open by the results we have obtained with our method
are especially interesting in this respect due to its simplicity and 
accuracy.

\vspace{.5cm}

\begin{table}
\caption{Values of $\sigma_{\theta,{\rm i}}^2$ and $T(E)$
obtained with our model, mean beam energies at the isocenter
$\overline{E}_0$ (see text) and 
$T(E)$ values found by interpolating those given in Table~2.6 of 
Ref.~\protect\cite{icr84} for these mean energies and corrected for 
M{\o}ller scattering.}
\begin{tabular}{ccccccccc}
 & & \multicolumn{3}{c}{``Our model''} & & & &
Ref.~\protect\cite{icr84} \\
\cline{3-5} \cline{9-9}
\rule{0mm}{4mm}Nominal Energy &~~& $\sigma_{\theta,{\rm i}}^2$ &~~&
 $T(E)$ &~~& $\overline{E}_0$ &~~& $T(\overline{E}_0)$ \\[.2mm]
\protect[MeV] && [rad$^2$] && [rad$^2$~cm$^{-1}$] && [MeV] &&
[rad$^2$~cm$^{-1}$] \\ \hline
~6 && $ (25.5 \pm 0.6)\times 10^{-4} $ && 
$ (12.5 \pm 0.6)\times 10^{-5}$ & & ~5.3 & & $24.983 \times 10^{-5}$ \\
~8 && $ (16.2 \pm 0.4)\times 10^{-4} $ && 
$ (~8.8 \pm 0.3)\times 10^{-5}$ & & ~7.2 & & $14.663 \times 10^{-5}$ \\
12 && $ (10.1 \pm 0.2)\times 10^{-4} $ && 
$ (~3.3 \pm 0.2)\times 10^{-5}$ & & 11.0 & & $~7.015 \times 10^{-5}$ \\
15 && $ (~8.9 \pm 0.2)\times 10^{-4} $ && 
$ (~1.6 \pm 0.2)\times 10^{-5}$ & & 13.9 & & $~4.670 \times 10^{-5}$\\
18 && $ (~6.2 \pm 0.2)\times 10^{-4} $ && 
$ (~0.6 \pm 0.1)\times 10^{-5}$ & & 16.9 & & $~3.324 \times 10^{-5}$\\
\end{tabular}
\end{table}

\section{Summary and conclusions}

In this work we have proposed, developed and tested a new
procedure to determine the spatial spread of clinical electron beams.
Two are the main advantages of the method proposed. First, we have
assumed that the dose profiles can be described by means of a function
which includes both a Gaussian part and a background function, this
last taking into account those processes not considered in the first
one. Second, this new method is based on the direct fit of the
dose profiles measured at the center of the beam and below a lead block
covering half of the beam. The beam divergence is incorporated to 
the model in a very easy way and its consequences (the inverse squared
law and the linear shift of the centroid) are checked to be fulfilled
with a high accuracy. Besides, the fitting procedure provides the 
spatial spread in a straightforward way and a great part of the 
ambiguities and errors of the usual methods based in penumbra 
measurements are eliminated. 

The data obtained in this way for the spatial spread have been fitted 
to a quadratic-cubic function of the distance $z$. The nice
fits obtained confirm the behavior expected from the Fermi-Eyges 
theory for the Gaussian part of our model and open the
possibility for using our approach to measure the scattering power in
air. 

We think that the model we propose in this work deserves a more
detailed study because of its possible application to describe the
clinical electron beam behavior. Obviously, it is necessary to apply
it to a variety of different measuring circumstances in order to complete
its validity for this purpose. Thus, it is mandatory to proceed with
the applicator, what will provide us information clinically relevant.
Also, the results obtained with phantoms of, e.g., water will permit to
elucidate the accuracy of our procedure for more practical situations.
Work in these directions is in progress.

\acknowledgements{We gratefully acknowledge useful comments and discussions
with Dr. P. Gal\'an. We also are indebted to the direction of the 
Hospital Universitario ``San Cecilio'' for permitting us the use of the
accelerator to carry the measurements. This work has been supported in
part by the Junta de Anadaluc\'{\i}a (Spain).}

\newpage

\vspace{1cm}

\vspace{1cm}

\vspace{1cm}

\newpage

\end{document}